\documentclass[12pt, letterpaper]{article}
\usepackage[utf8]{inputenc}
\usepackage{amsmath}
\usepackage{mathtools}
\usepackage{graphicx}
\usepackage{bbding}
\usepackage{float}
\usepackage{array}
\usepackage{ragged2e}
\usepackage{rotating}
\usepackage{url}
\usepackage[english]{babel}
\usepackage{csquotes}
\usepackage{lineno}
\usepackage{authblk}
\usepackage{amsfonts}
\usepackage{amsthm,amssymb}
\usepackage{caption}
\usepackage{subcaption}
\usepackage{booktabs}
\usepackage{xr}
\usepackage[T1]{fontenc}
\usepackage{newtxmath,newtxtext}
 \usepackage[margin=1in]{geometry}
\usepackage[style=authoryear, maxcitenames=2, uniquelist=false, maxbibnames=10]{biblatex}

\makeatletter
\renewcommand{\fnum@figure}{Fig. \thefigure}
\makeatother
\newcolumntype{x}[1]{>{\centering\arraybackslash\hspace{0pt}}p{#1}}

\DeclarePairedDelimiter{\norm}{\lVert}{\rVert}

\providecommand{\keywords}[1]
{
  \small	
  \textbf{Keywords---} #1
}

\title{Bayesian Multi-Species N-Mixture Models for Unmarked Animal Communities}

\author[1,2]{Niamh Mimnagh \thanks{Correspondence author. E-mail: niamh.mimnagh.2013@mumail.ie}}
\author[1,2,3]{Andrew Parnell}
\author[1,2,3]{Estev\~{a}o Prado}
\author[1,2]{Rafael de Andrade Moral}
\affil[1]{\small Hamilton Institute, Maynooth University, Maynooth, Co. Kildare, Ireland}
\affil[2]{\small Department of Mathematics and Statistics, Maynooth University, Maynooth, Co. Kildare, Ireland}
\affil[3]{\small Insight Centre for Data Analytics, Maynooth University, Maynooth, Co. Kildare, Ireland}
\date{}   

\begin{document}

\maketitle

\vspace{-1.8cm}
\abstract{We propose an extension of the N-mixture model that enables the estimation of abundances of multiple species as well as the correlations between them. Our novel multi-species N-Mixture model (MNM) is the first to address the estimation of both positive and negative inter-species correlations, which allows us to assess the influence of the abundance of one species on another. We provide extensions that permit the analysis of data with excess of zero counts, and relax the assumption that populations are closed through the incorporation of an autoregressive term in the abundance. Our approach provides a method of quantifying the strength of association between species’ population sizes and is of practical use to population and conservation ecologists. We evaluate the performance of the proposed models through simulation experiments in order to examine the accuracy of both model estimates and coverage rates. The results show that the MNM models produce accurate estimates of abundance, inter-species correlations and detection probabilities at a range of sample sizes. The MNM models are applied to avian point data collected as part of the North American Breeding Bird Survey (NABBS) between 2010 and 2019. The results reveal an increase in Bald Eagle abundance in south-eastern Alaska in the decade examined.}


\section{Introduction}
Abundance in animal communities is of great interest in ecology, particularly in the areas of conservation and wildlife management \parencite{witmer2005wildlife, nichols2004abundance}. Count data is an attractive option for estimating abundance due to the relative affordability with which  it may be collected and the reduced risk of harm to both animals and humans when compared to more direct data collection methods \parencite{verdade2013counting}. However, count data for animal abundance has a tendency to suffer from imperfect detection (i.e., the recorded information is usually imperfect in the sense that it does not represent the total abundance). Furthermore, when the detection probability is small, there is a tendency towards the underestimation of abundance. Due to the characteristics of these data, traditional modelling techniques, such as generalised linear models \parencite{mccullagh1989generalized}, cannot be applied directly to the data, as they do not accommodate imperfect detection.

N-mixture models \parencite{RoyleJ2004} constitute a class of models which may be used to estimate abundance from count data. These models assume that the population under analysis is closed, i.e., it is constant in terms of births, deaths, and migration. The counts at each site and time are considered independent and identically distributed (i.i.d) random variables that follow a Binomial distribution. In the original N-mixture model, the detection probability is estimated using the data, without the specification of any prior distribution with fixed parameters. The population size at each site is treated as a random effect, with an assumed probability distribution. The distributions that are typically considered for the population size at each site are the Poisson and Negative Binomial, although any other non-negative discrete distribution could also be considered. 

The ability to estimate correlations between species abundances allows us to relax any assumption of independent species abundances. This is the aim of the multi-species N-mixture (MNM) models presented in this paper -- a new class of models which estimate abundance for multiple species simultaneously while accounting for imperfect detection, and estimate inter-species correlations, which are intended to allow for inferences about the relationships between different species.

The remainder of the paper is organised as follows. In Section~\ref{sec:methods}, we introduce our novel modelling framework to estimate abundance and inter-species correlations in animal communities based on spatio-temporal count data. We also describe the model formulation, estimation procedure, and the computation of the inter-species correlations. In Section~\ref{realworldapplication}, we present the data obtained from the North American Breeding Bird Survey (NABBS) \parencite{Pardieck2020}, which will be used to illustrate our modelling approach. Later, in Section~\ref{sec:results}, we compare results of model fit on the NABBS data to obtain the best fit. Finally, in Section~\ref{sec:discussion}, we present a general discussion.

\section{Related Works}
Several multi-species modelling frameworks have been developed previously which allow for the analysis of occurrence-data \parencite{dorazio2005estimating, yamaura2011modelling} or count-data \parencite{yamaura2012biodiversity, golding2017multispecies, Gomez2017} of multiple species. 

\textcite{dorazio2005estimating} developed a model for estimating the size of a biological community by modelling the probability of detection as a Binomial random variable, and the probability of occurrence as a Bernoulli random variable. They allow rates of detection and occurrence to vary among species, and not every species is assumed to be present at every location. However, the aim of their model is to determine the number of species, not the number of individuals of each species, as is the aim of N-mixture models.

\textcite{yamaura2011modelling} developed a multi-species model that estimates the animal abundance from occurrence-data. This is an extension of the single-species model developed by \textcite{royle2003estimating}, in which binary detection/non-detection data is linked to abundance. \textcite{yamaura2012biodiversity} extended this model to count data. The assumption behind these models is that the abundances or detection probabilities of species in the community might be linked by species-level or functional group-level characteristics. However, inter-species abundance correlations are not explored within these models.

\textcite{Gomez2017} developed a multi-species N-mixture model whose aim was to allow for the estimation of abundance of rare species by borrowing strength from other species in the community. This was done by assuming detection probabilities are drawn at random from a Beta distribution. Another multi-species N-mixture model was developed by \textcite{golding2017multispecies}, which used the dependent double-observer method to create a multi-species dependent double-observer abundance model. This allowed them to address an issue of false-positive errors in detection. The focus of both \textcite{Gomez2017} and \textcite{golding2017multispecies} was an improvement in detection probability. None of the preceding multi-species models allow us to make inferences as to the relationships within an ecological community, as we propose to do with our multi-species N-mixture model.

\textcite{moral2018models} developed an extension to the single-species N-mixture model which allowed for the estimation of abundances of two species, and the correlations between these abundances. However, this model only examines two species, and is therefore not as complete as the model we propose here, which allows us to examine whole communities.

\textcite{dorazio2014estimating} developed a multi-species N-mixture model which allowed for abundances of species with similar traits to be correlated. However, to guarantee positive definite correlations, they only allow for positive correlations through the use of a distance metric $d$ coupled with a spatial autocorrelation structure of the type $e^{-\frac{d}{\phi}}$. The framework we present here is more complete in that we guarantee positive definiteness of the correlation matrix via an elegant prior setup. We also explore ways of incorporating zero-inflation and open population dynamics, which is not something attempted by \textcite{dorazio2014estimating}. 

Finally, \textcite{niku2019gllvm} describe generalised linear latent variable models - a modelling technique which allows for obtaining correlation matrices in an elegant manner. However, these models do not allow for the incorporation of imperfect detection.

\section{Methods}
\label{sec:methods}

The models developed in the following Section are a multi-species extension to the original N-mixture model of \textcite{RoyleJ2004}, which allows for accurate estimation of both the latent abundances and inter-species correlations, while accounting for imperfect detection and relaxing the closure assumption.

\subsection{Multi-Species N-Mixture Model (MNM Model)}\label{MNMmodel}
Consider a study which sees count data $Y_{its}$ collected, where $Y_{its}$ is the number of individuals observed for $S$ different species ($s=1,\ldots,S$) from $R$ sites ($i=1,\ldots,R$). Consider also that these samples are taken from each site on $T$ occasions ($t=1,\ldots,T$). The true abundance at site $i$ for species $s$ is given by $N_{is}$. We observe $N_{is}$ with detection probability $p_{its}$, and it is assumed that species populations are closed with respect to births, deaths and migration (i.e., that the population sizes do not change due to any of these factors, akin to the N-mixture model proposed by \textcite{RoyleJ2004}). Our model assumes that $N_{is}$ follows a Poisson distribution, and may be written as:
\begin{linenomath*}
\postdisplaypenalty=0
\begin{align}
    Y_{its} \mid N_{is}, p_{its} & \sim \text{Binomial}(N_{is}, p_{its}),\nonumber\\
    N_{is} \mid \lambda_{is} & \sim \text{Poisson}(\lambda_{is}), \nonumber \\
    \mbox{logit($p_{its}$)} &=  \textbf{z}^{\top}_{it} \textbf{b}_{s}, \nonumber \\
    \mbox{log($\lambda_{is}$)}  &= a_{is} + \textbf{x}^{\top}_{i}\boldsymbol\beta_{s}, \nonumber \\
    \mathbf{a}_{i} \mid \boldsymbol\mu_{a}, \boldsymbol{\Sigma}_{a} & \sim \mbox{MVN}(\boldsymbol{\mu}_{a}, \boldsymbol{\Sigma}_{a}), \nonumber
\end{align}
\end{linenomath*}
where $\textbf{a}_{i} = (a_{i1}, \ldots, a_{iS})^{\top}$. The Poisson rate parameter $\lambda_{is}$ represents the mean abundance at site $i$, and $\mathbf{a}_{i}$ is an $S$-dimensional vector that contains the random effects $a_{is}$ that allow us to estimate inter-species correlations. In the above model, covariates may be incorporated in the detection probability and the abundance, with $\mathbf{z}^\top_{it}$ the $it$-th row of the design matrix $\mathbf{Z}$ of dimension $RT \times q_p$, $b_{s}$ the $q_{p} \times 1$ parameter vector for the probability of detection, $\mathbf{x}^\top_{i}$ the $i$-th row of the design matrix $\mathbf{X}$ of dimension $R \times q_{\lambda}$, and $\boldsymbol{\beta}$ the $q_{\lambda} \times 1$ parameter vector for the abundance. Here, $q_{p}$ and $q_{\lambda}$ represent the number of covariates associated with the detection probability and the abundance, respectively. Note that different covariate effects may be estimated per species, and other species-level random effects may also be included.

\subsection{Hurdle-Poisson Model (MNM-Hurdle Model)}
In this Section, we develop a further extension of the multi-species N-mixture model, appropriate for scenarios in which the number of zero-counts exceed those expected under a Poisson distribution. We now allow the counts to follow a Hurdle-Poisson distribution, with $\lambda_{is}$ defined as in the MNM Model, and $\theta$ the probability of obtaining a zero-count. 

The Hurdle-Poisson distribution consists of two separate processes. The first is a Bernoulli process, which determines whether a site is occupied (count is non-zero) or unoccupied (count is zero). If the count is non-zero, a second random variable with a zero-truncated Poisson distribution determines the value of the count, i.e.,
\begin{linenomath*}
\postdisplaypenalty=0
\begin{align}
&\text{Occupancy}_{is}  \sim \text{Bernoulli}(1-\theta), \nonumber \\
&\text{Count}_{is} \sim \text{Zero Truncated Poisson}(\lambda_{is}). \nonumber
\end{align}
\end{linenomath*}
We then define the latent abundances $N_{is}$ as
\begin{linenomath*}
\[
  N_{is} = \left.
  \begin{cases}
    0 & \text{if } \text{Occupancy}_{is} = 0 \\
    \text{Count}_{is} & \text{if } \text{Occupancy}_{is} = 1
  \end{cases}
  \right. ,
\]
\end{linenomath*}
which yields
\begin{linenomath*}
$$N_{is}\sim\text{Hurdle-Poisson}(\lambda_{is},\theta).$$
\end{linenomath*}
\noindent
If the Bernoulli process is equal to 0, then the site is unoccupied and $N_{is}$ is equal to 0. However, if the Bernoulli process is equal to 1, then the hurdle is crossed, and the value of $N_{is}$ is determined by the zero-truncated Poisson process. Similar to the MNM model, populations are assumed to be closed.

We assume a single probability of obtaining a zero count $\theta$. However, $\theta$ may also be allowed to vary by site and/or species, and may depend on covariates through a logit link. All other parameters are distributed as described in the MNM model in Section \ref{MNMmodel}.

\subsection{Autoregressive Model (MNM-AR Model)}
In order to model populations over multiple years, a further extension to the multi-species N-mixture model is proposed, which allows us to relax the assumption that species populations are closed with respect to births, deaths and migration. We do this through the inclusion of an autoregressive term in the abundance parameter.

The study design now consists of data collected over $K$ years ($k=1,\ldots,K$) for $S$ species at $R$ locations, each with $T$ sampling occasions. 
The observed abundance ($Y$) and actual abundance ($N$) are now allowed to vary by year:
\begin{linenomath*}
\postdisplaypenalty=0
\begin{align*}
    &Y_{itks} \sim \text{Binomial}(N_{iks}, p_{s}),\\
       & N_{iks} \sim \text{Poisson}(\lambda_{iks}).
\end{align*}
\end{linenomath*}
If $k=1$, then $\lambda_{i1s}$ is defined as before:
\begin{linenomath*}
\begin{equation*}
  \text{log}(\lambda_{i1s}) = a_{is} + \textbf{x}^{\top}_{i}\boldsymbol\beta_{s}. \nonumber \\
\end{equation*}
\end{linenomath*}
However, for $k>1$, we allow $\lambda_{iks}$ to depend on the latent abundance at year $k-1$:
\begin{linenomath*}
\begin{equation*}
  \text{log}(\lambda_{iks}) = a_{is} + \textbf{x}^{\top}_{i}\boldsymbol\beta_{s} + \phi_{s}\text{log}(N_{i(k-1)s}+1).
\end{equation*}
\end{linenomath*}
The term $\text{log}(N_{i(k-1)s}+1)$ is used here rather than the simpler $N_{i(k-1)s}$ to avoid the rapid increase in sampled $\lambda$ values when $N$ values are large \parencite{Foikanos2011}.

\subsection{Hurdle-Autoregressive Model (MNM-Hurdle-AR Model)}

A straightforward combination of the MNM-Hurdle model and the MNM-AR model produces the MNM-Hurdle-AR model. This model accommodates excess zeros while also accounting for an autoregressive structure in the data. The zero-inflation is introduced as in the MNM-Hurdle model, i.e.,
\begin{linenomath*}
\postdisplaypenalty=0
\begin{align}
    Y_{itks} & \sim \text{Binomial}(N_{iks}, p_{s}), \nonumber \\
        N_{iks} & \sim \text{Hurdle-Poisson}(\lambda_{iks}, \theta), \nonumber
\end{align}
\end{linenomath*}
where
\begin{linenomath*}
\[
  \text{log}(\lambda_{iks}) = \left.
  \begin{cases}
    a_{is} + \textbf{x}^{\top}_{i}\boldsymbol\beta_{s}, & \text{for } k = 1 \\
    a_{is} + \textbf{x}^{\top}_{i}\boldsymbol\beta_{s}  + \phi_{s}\text{log}(N_{i(k-1)s}+1),\nonumber & \text{for } k>1
  \end{cases}
  \right. .
\]
\end{linenomath*}
\noindent

\subsection{Model Estimation}
\label{modelEstimation}
The models described in this paper are implemented using a Bayesian framework. Each of the above models were implemented in R \parencite{Rsoftware} through the probabilistic programming software JAGS \parencite{plummer2003jags, Plummer2017} using four chains with 50,000 iterations each, of which the first 10,000 were discarded as burn-in, and a thinning of five to reduce autocorrelation in the MCMC samples. Parameter convergence was determined using the potential scale reduction factor ($\hat{R}$), a diagnostic criteria proposed by \textcite{gelman1992inference}. An $\hat{R}$ value that is very close to one is an indication that the four chains have mixed well. If $\hat{R}$ value was less than 1.05, the chains were considered to have mixed properly, and the posterior estimates of the parameters were considered reliable.

Prior distributions were assigned as follows: $\boldsymbol{\mu}_{a}$, the vector of means of the random effect $\mathbf{a}$, was assigned a multivariate Normal prior with a diagonal variance-covariance matrix $\boldsymbol{\Sigma}_{0}$ and mean vector $\boldsymbol{\mu}_{0}$. $\boldsymbol{\Sigma}_{a}$, the variance-covariance matrix of $\mathbf{a}$ was assigned an inverse-Wishart prior with a diagonal scale matrix $\boldsymbol{\Omega}$, and $S+1$ degrees of freedom $v$ which results in a Uniform(-1,1) prior on the correlations \parencite{Plummer2017}:
\begin{linenomath*}
\postdisplaypenalty=0
\begin{align}
        \boldsymbol{\mu}_{a} & \sim \text{Normal}_{s}(\boldsymbol{\mu}_{0},\boldsymbol{\Sigma}_{0}), \nonumber \\
        \boldsymbol{\Sigma}_{a} & \sim \text{Inverse-Wishart}(\boldsymbol{\Omega}, v). \nonumber 
\end{align}
\end{linenomath*}

An inverse-Wishart distribution is specified as the prior for the covariance matrix of the random effect $\textbf{a}$. Criticisms of the inverse-Wishart prior include the dependency imposed between correlations and variances, and the fact that there is a single degree of freedom parameter which determines the uncertainty for all variance parameters. It is demonstrated by \textcite{alvarez2014bayesian} that when the variance is small relative to the mean, the correlation is biased towards zero, and the variance is biased towards larger values, though when working with count data, typically variances are large relative to the mean. Despite these issues, the inverse-Wishart distribution is a prior distribution commonly assigned to a covariance matrix in Bayesian analysis due to its conjugacy with the Normal distribution, and for these models the inverse-Wishart distribution provides a good solution due to its guarantee of providing a positive definite covariance matrix.

In the Hurdle and Hurdle-AR models, $\theta$ is assigned a Beta prior with the value of both shape parameters equal to one, which is equivalent to a vague uniform prior:
\begin{linenomath*}
\postdisplaypenalty=0
\begin{align}
       \theta \sim \text{Beta}(1, 1). \nonumber
\end{align}
\end{linenomath*}
In the AR and Hurdle-AR models, $\phi$ is assigned a Multivariate Normal prior, with hyperpriors $\boldsymbol{\mu_{\phi}}$ and diagonal matrix $\boldsymbol{\Sigma_{\phi}}$:
\begin{linenomath*}
\postdisplaypenalty=0
\begin{align*}
    \boldsymbol{\phi_{s}} & \sim \text{MVN}(\boldsymbol{\mu_{\phi}}, \boldsymbol{\Sigma_{\phi}}), \nonumber
\end{align*}
\end{linenomath*}

Extensive simulation studies were carried out to examine the accuracy of parameter estimates; see Appendix \ref{AppendixB} for more details.

\subsection{Inter-Species Correlations}
The presence of the multivariate normal random effect $\textbf{a}$ in the abundance provides a link between species’ abundances. The correlation matrix for the random effect, $\boldsymbol{\Sigma}_{a}$, may be estimated directly from the Bayesian model. In this sense, the inter-species correlations for the latent abundances $N_{s}$ and $N_{s'}$, for all $s \neq s'$, are calculated for each model as:
\begin{linenomath*}
\postdisplaypenalty=0
\begin{align}
    \rho(N_{s}, N_{s'})= \frac{\text{Cov}(N_{s}, N_{s'})}{\sqrt{(\text{Var}(N_{s}))(\text{Var}(N_{s'}))}}.\nonumber
\end{align}
\end{linenomath*}
The derivation of $\text{Cov}(N_{s}, N_{s'})$ can be found in Appendix \ref{AppendixA}.

The inter-species correlations for the MNM model and Hurdle model are assumed not to vary by year, so these models have a single analytic correlation matrix. However, in the AR and Hurdle-AR model, we assume latent abundances change by year, which requires the computation of $K$ analytic correlation matrices. Note that the MNM and AR models required the use of properties of conditional variance and covariance to determine analytic correlations. In the Hurdle and Hurdle-AR models, the properties of conditional variance and covariance were merged with second-order Taylor approximations to make their computation feasible.

\section{Case Study: North American Breeding Bird Survey} \label{realworldapplication}
In this section, we describe the application of the multi-species N-mixture models to a real world case study, to examine bird populations using data collected as part of the North American Breeding Bird Survey (NABBS).
 
The North American Breeding Bird Survey \parencite{Pardieck2020} was first conducted in 1966, and now provides data annually on more than 400 bird species across 3700 routes in the United States and Canada. Each of these routes is approximately 24.5 miles long and is composed of 50 stops, approximately 0.5 miles apart. At each stop, every bird seen or heard within a 0.25-mile radius is recorded. For the sake of our models, each of these routes is considered a site, and each of the 50 stops along a route is a sampling occasion. 

\begin{figure}[]
    \centering
    \includegraphics[width=0.65\textwidth]{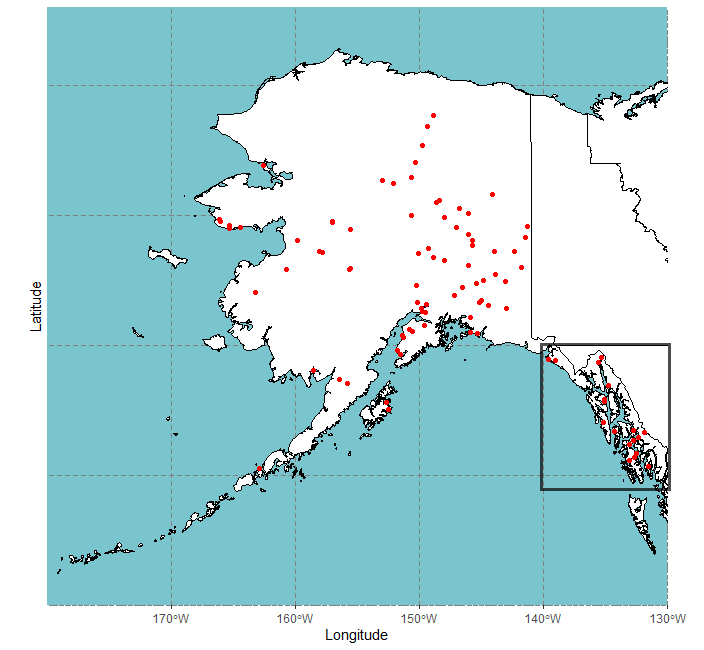}
    \caption{Location of sites in Alaska marked in red, with the Alexander Archipelago - the location of the Alaskan Bald Eagle population - outlined in black.}
    \label{fig:Figure1}
\end{figure}

We examine data collected in Alaska in the 10-year period 2010-2019. There are 94 routes in Alaska (Fig.~\ref{fig:Figure1}) at which data was collected during this time, and each of these routes are composed of 50 sampling locations, totalling 4,700 observations per bird species.

Bald Eagle populations in Alaska are estimated at between 8,000 and 30,000 birds, accounting for roughly half of the global population \parencite{hodges2011bald, hansen1987regulation, king1972census}. For this reason, Bald Eagles were chosen as a species of interest. Several other species were chosen; these included waterbirds such as geese, swans and snipes which were chosen for their relationships with Bald Eagles, as Bald Eagles are known to prey on waterbirds such as ducks, geese and grebes when fish are in short supply \parencite{dunstan1975food, todd1982food, mcewan1980food}. Additionally, a selection of species with inland habitats, such as thrushes and swallows, were examined. In total, 10 species were selected for analysis, of the 233 total species present in Alaska within the 10-year period. The full list of species selected and the frequency with which they were observed is given in Table \ref{table:frequency}.

The models described in Section~\ref{sec:methods} were fitted to the NABBS data. Each was fitted three times, varying the dimension of the detection probability. Initially, detection probability was allowed to vary by site, species and year. Subsequently, models were fitted in which detection probability varies only by site and species, and then by species alone.

\begin{table}
\begin{tabular}{x{4.3cm} *{10}{x{0.6cm}}}
\hline
   & 2010 & 2011 & 2012&2013 & 2014& 2015 & 2016 & 2017 & 2018 & 2019 \\
 \hline
Bald Eagle & 21 & 20 & 25  & 20 & 22 & 19 & 23 & 30 & 30 & 23\\
  Canada Goose & 14 & 18 & 17 & 19 & 8 & 11 & 12 & 15 & 16 & 14\\
  Hammond's Flycatcher&16&16 & 15& 12& 12& 11& 12& 15& 17& 14\\
  Red-breasted Sapsucker & 12& 11& 10& 9& 10& 9& 10& 12& 12& 12\\
  Steller's Jay &16 &  16 &14&11&12&13&12&14&15 &13\\
  Swainson's Thrush&57  & 52&56&50 &50 &49 &48&63 &62& 55\\
  Tree Swallow &27   &26&27&25&22&27&25&31&27&30\\
  Trumpeter Swan &13   & 9 &14&14&11&12&12&14 &12&8\\
  Varied Thrush &62 &  58 &62&52&55& 52& 50& 68&62 &58\\
  Wilson's Snipe&52  & 47&  51&47&44&48 &42&53 &50&47\\
 \hline
\end{tabular}
\caption{Number of routes each species appears at by year, out of a total 94 routes.}
\label{table:frequency}
\end{table}

Initially, the models were fitted without covariates, and results were compared using their Bayesian Information Criterion (BIC) \parencite{delattre2014note} values. Subsequently, latitude, longitude and their interaction term latitude $\times$ longitude were included in the linear predictors for the abundance parameters, and models were again compared using BIC values. All covariates were scaled to have zero-mean and unit variance. 

Initial examination of this data revealed that $93.2\%$ of observations ($438,040$ of a total of $470,000$ observations) consisted of zero counts. This suggested that a model with a hurdle component might provide an appropriate framework for this data. Furthermore, this data was collected over the course of a decade. For this reason, we might expect that an autoregressive term may be useful to incorporate the time dependence. 

Each model was fitted using four chains with 50,000 iterations each, of which the first 10,000 were discarded as burn-in, using a thinning value of five. All prior distributions were assigned as described in Section \ref{modelEstimation}.

\section{Results}
\label{sec:results}
Initially, the models were fitted without covariates and were compared using BIC values. The result of this comparison was that the Hurdle-AR model, in which detection probability varies by species (Hurdle-AR(C)), provided the best fit for the NABBS data. However, the addition of a response surface for latitude and longitude in the linear predictors for the abundance parameters results in the Hurdle model in which detection probability varies by species (Hurdle(C)) producing the lowest BIC value. This suggests that, within the range of models produced, this model provides the best fit for our data. The variance which was initially explained by the addition of the autoregressive term is now explained by the latitude and longitude covariates, which render the autoregressive component unnecessary. The result of this comparison is given in Table \ref{tab:ABBSBIC}. 

{\footnotesize
\renewcommand{\baselinestretch}{1.2}
\begin{table}[H]
    \centering
    \begin{tabular}{lrrrr}
    \hline 
    & \multicolumn{2}{c}{No Covariates} & \multicolumn{2}{c}{Covariates}\\
    \multicolumn{1}{c}{Model}  & \multicolumn{1}{c}{Parameters} & \multicolumn{1}{c}{BIC} & \multicolumn{1}{c}{Parameters} & \multicolumn{1}{c}{BIC} \\

        \hline
    MNM(A) &9,465 & 291,860 & 9,471 &291,899\\
    MNM(B) &1,005 & 194,592&1,011 &194,743\\
    MNM(C) &75 &183,587 &81 &	183,940\\
    MNM-AR(A) &9,467 & 291,648 &9,473 &291,808\\
    MNM-AR(B)&1,007 & 194,767 &1,013 &194,805\\
    MNM-AR(C)&77 & 183,922 &83& 183,540\\
    MNM-Hurdle(A) &9,466 & 291,699 &9,472 &291,596\\
    MNM-Hurdle(B) &1,006 & 194,964 &1,012 &194,915\\
    MNM-Hurdle(C) &76 & 183,890 &82 & \textbf{183,349}\\
    MNM-Hurdle-AR(A) &9,468 & 291,597 &9,474 &291,652\\
    MNM-Hurdle-AR(B) & 1,008& 194,859 &1,014&	195,044\\
    MNM-Hurdle-AR(C) &78 & \textbf{183,572} &84 & 183,644\\
        \hline
    \end{tabular}
    \caption{Number of estimated parameters and BIC values comparing model fits on NABBS data. (A) contains detection probability which varies by site, species and year, (B) contains detection probability which varies by site and species, (C) contains detection probability which varies only by species. Smallest BIC values for each case are indicated in bold.}
    \label{tab:ABBSBIC}
\end{table}
}

The latent inter-species correlations are given in Fig.~\ref{fig:Figure3}, while the derivation of analytic correlations, which vary by site and year, are given in Appendix \ref{AppendixA}. The latent correlations are obtained after probability of detection and other covariates are taken into account. They may be interpreted as an interaction strength metric, which allows for the study of the influence of one species' abundance on the others \parencite{berlow2004,moral2018models}.

\begin{figure}
    \centering
    \includegraphics[width=0.75\textwidth]{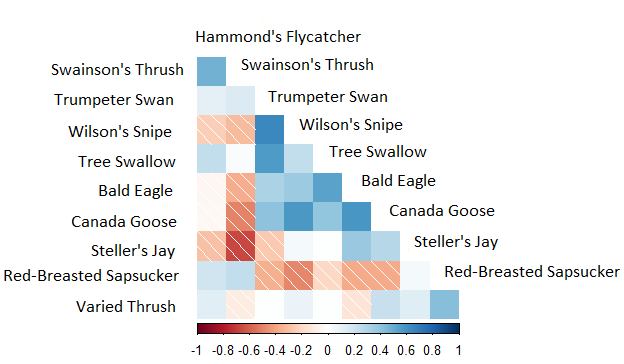}
    \caption{Estimated latent inter-species correlation matrix, produced by the Hurdle(C) model fitted to the NABBS data including covariates in the linear predictor for the abundance parameter.}
    \label{fig:Figure3}
\end{figure}

\section{Discussion}
\label{sec:discussion}

We have proposed a multi-species extension to the N-mixture model which allows for the estimation of inter-species abundance correlations through the addition of a random variable in the abundance. Results of simulation studies (see Appendix \ref{AppendixB}) reveal that this model performs well under a range of scenarios, with abundances and detection probabilities that range from low to high. For this reason, we believe that this approach represents an attractive framework for examining multi-species abundances.

Issues with parameter convergence were encountered when fitting the Hurdle and Hurdle-AR models. When zero-inflation and abundance are large, and detection probability is small, issues with convergence occurred in up to 20$\%$ of parameters. While this convergence issue does not appear to negatively affect the relative biases of parameter estimates (as can be seen in Appendix \ref{AppendixB}, Table \ref{table:small} and Table \ref{table:large}), coverage probability for detection probability $p$ and random effect mean $\mu_{a}$ is negatively impacted (Appendix \ref{AppendixD}). In the same models, we see larger coverage for $N$. This is to be expected, and is due to zero counts being perfectly predicted.

Previous works have demonstrated that N-mixture models can sometimes suffer from issues with identifiability \parencite{dennis2015computational} wherein probability of detection estimates are very close to zero and abundance estimates are infinite. To address this issue, we have performed extensive simulation studies, detailed in Appendix \ref{AppendixB}, in which we assess the estimates of abundance and detection probability for a large range of sample sizes, detection probabilities, abundance sizes, and in the case of the Hurdle and Hurdle-AR models, zero-count probabilities. The result was a simulation study which demonstrated no evidence that this modelling framework suffers from these identifiability issues. 

The models presented here all use the Poisson distribution to model the latent abundances. However, any other count distribution might instead be used, for example, the Negative Binomial. Our calculations for the analytic correlations, however, reflect only the use of the Poisson distribution.

Case study results reveal that the difference in BIC values between the model with the lowest BIC value (Hurdle(B) with covariates) and the model with the second-lowest BIC value (Hurdle-AR(B) without covariates) is 223. This sizeable difference in BIC values suggests that the Hurdle(B) model with covariates provides a better fit than the Hurdle-AR(B) model without covariates.

Case study detection probability values range from 0.047 (Tree Swallow) to 0.564 (Swainson's Thrush). Estimates of the maximum latent abundance $N$ per species are provided in Appendix \ref{AppendixC}, which reveals that while N-mixture models occasionally suffer from identifiability issues as described above, this does not appear to be an issue for this case study. 

\begin{figure}
    \centering
    \includegraphics[width=0.8\textwidth]{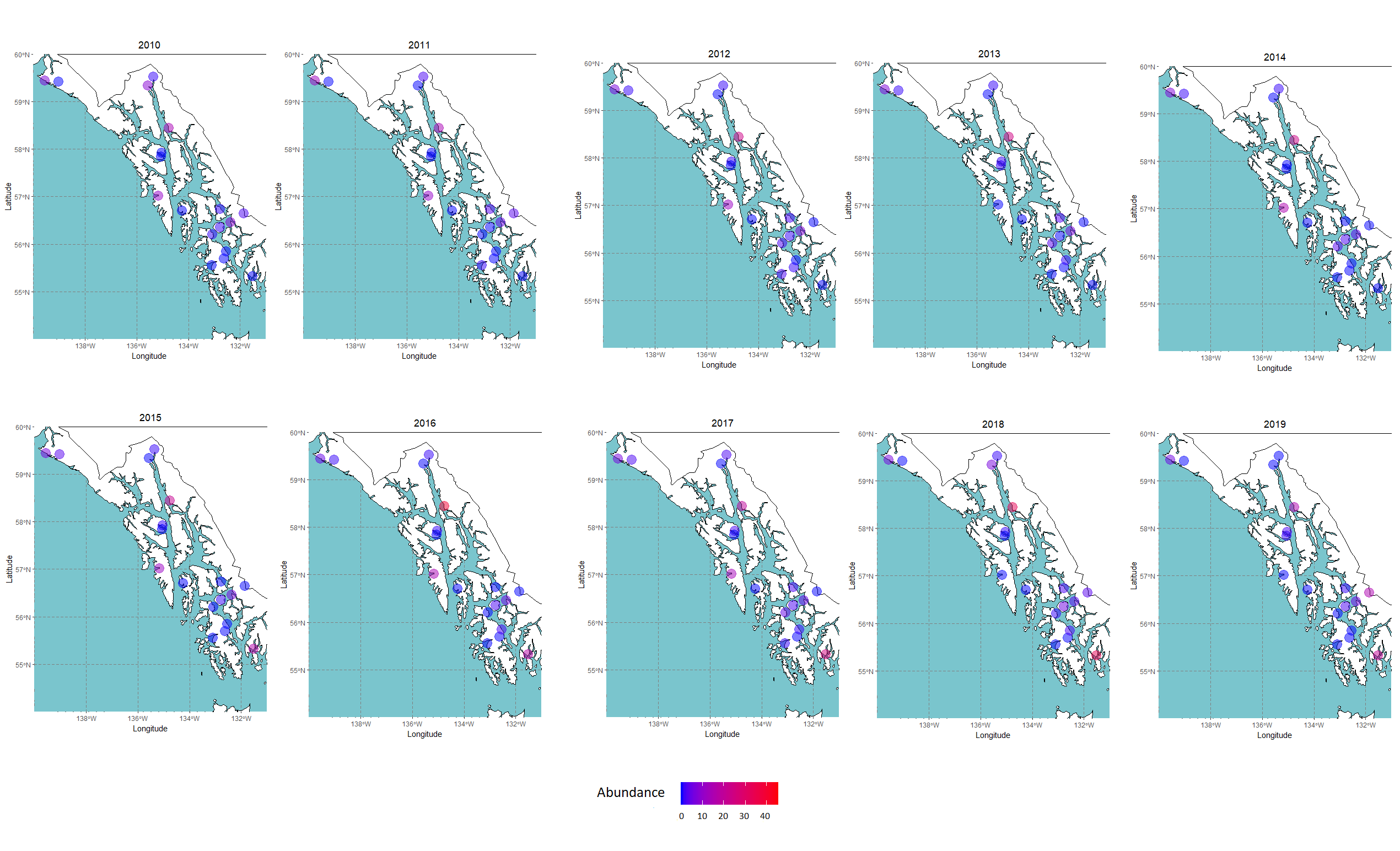}
    \caption{Estimated Bald Eagle abundances at sites in the Alexander Archipelago, produced by the Hurdle(B) model fitted to the NABBS data.}
    \label{fig:baldeagleabundance}
\end{figure}

\begin{figure}
    \centering
    \includegraphics[width=0.8\textwidth]{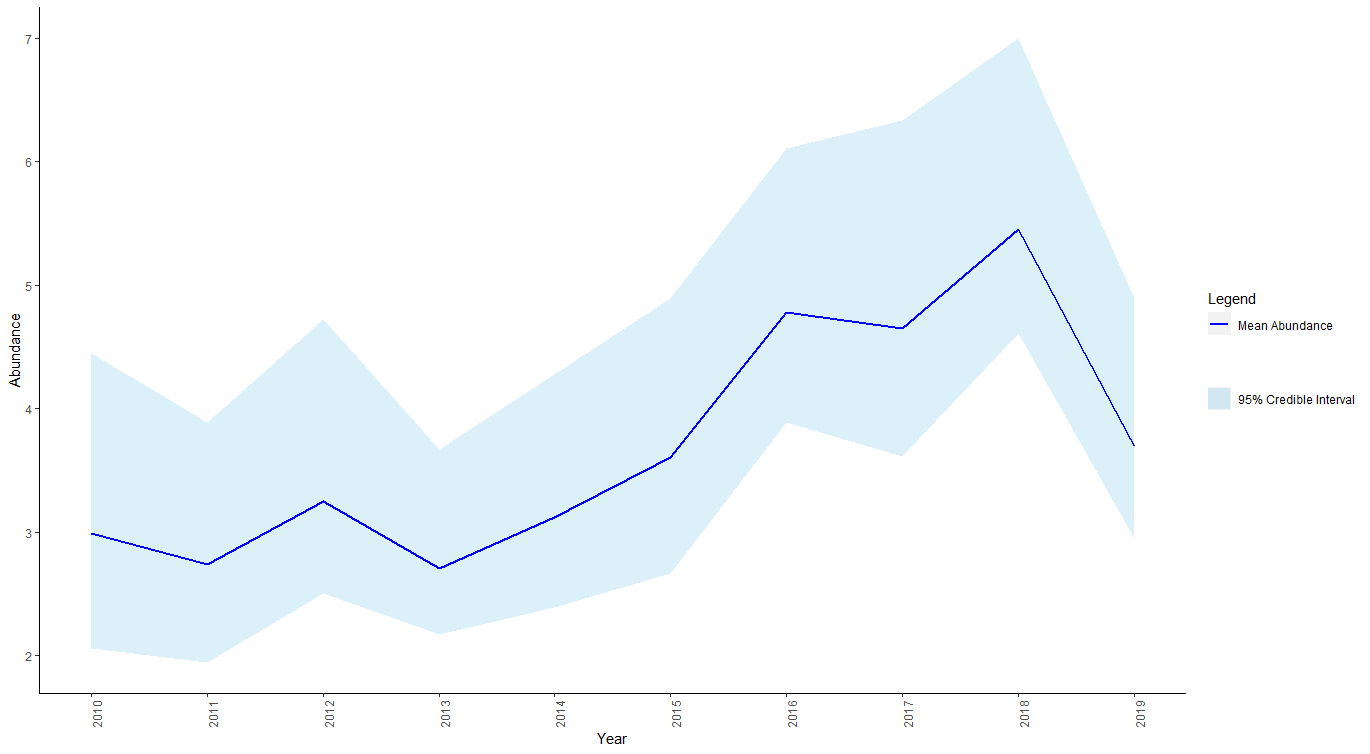}
    \caption{Annual mean abundance for Bald Eagles in the Alexander Archipelago, estimated by the Hurdle(B) model fitted to the NABBS data. The light-blue ribbon represents the 95\% credible interval for the mean.}
    \label{fig:baldEagleMeanAbundance}
\end{figure}

The estimates for Bald Eagle abundance produced by this model are plotted by site and year in Fig.~\ref{fig:baldeagleabundance}.
Of the 94 possible sites in Alaska, the Bald Eagle population is concentrated at 18 sites at the southeastern coast, along a 300-mile stretch of islands called the Alexander Archipelago. Examination of this figure suggested a possible increase in Bald Eagle abundance in this area between 2010 and 2019. The mean abundance was calculated per year (Fig.~\ref{fig:baldEagleMeanAbundance}), and a one-sided Mann-Kendall test \parencite{mann1945nonparametric, kendall1948rank} for an increasing trend in time series data was performed. The result of this was a Kendall's $\tau$ value of 0.6 and a p-value of 0.0082, indicating that it was appropriate to reject the null hypothesis that no increasing trend exists. We can therefore conclude that Bald Eagle abundances increased in the area of the Alexander Archipelago in the decade between 2010 and 2019.

In the models that contain an autoregressive component, we obtain separate $\text{Corr}(N_{s}, N_{s'})$ per year. As a feature of model formulation, the correlation between two species do not change sign from year to year. We can accommodate a change in sign by allowing for an unstructured covariance matrix of the autocorrelation coefficient $\boldsymbol{\Sigma_{\phi}}$, and this particular extension is subject of ongoing work. Furthermore, the models presented in this paper assume that sites are independent of one another. A further extension we are currently working on is the incorporation of spatial dependence.

\section{Declarations}

We are grateful to the associate editor and an anonymous referee, who helped to substantially improve the quality of the original manuscript.

\subsection{Funding}
Niamh Mimnagh's work was supported by a Science Foundation Ireland grant number 18/CRT/6049. The opinions, findings and conclusions or recommendations expressed in this material are those of the author(s) and do not necessarily reflect the views of the Science Foundation Ireland.

Estev\~{a}o Prado's work was supported by a Science Foundation Ireland Career Development Award grant number 17/CDA/4695 and SFI research centre 12/RC/2289P2. 

Andrew Parnell's work was supported by: a Science Foundation Ireland Career Development Award (17/CDA/4695); an investigator award (16/IA/4520); a Marine Research Programme funded by the Irish Government, co-financed by the European Regional Development Fund (Grant-Aid Agreement No. PBA/CC/18/01); European Union's Horizon 2020 research and innovation programme under grant agreement No 818144; SFI Centre for Research Training 18/CRT/6049, and SFI Research Centre awards 16/RC/3872 and 12/RC/2289P2.

\subsection{Competing Interests}
The authors have no competing interests to declare that are relevant to the content of this article.

\subsection{Code Availability}

Code for simulating data and fitting models is provided via the following link: \url{https://github.com/niamhmimnagh/MNM}.

\subsection{Data}
The North American Breeding Bird Survey data which was utilized for this research are as follows: \cite{Pardieck2020}, [available at \url{https://www.sciencebase.gov/catalog/item/52b1dfa8e4b0d9b325230cd9}]

\subsection{Ethics Approval}
Not Applicable.

\subsection{Consent}
Not Applicable.

\subsection{Author's Contributions}
All authors contributed to methodology design. Niamh Mimnagh and Rafael Moral analysed data and led the writing of the manuscript. Andrew Parnell and Estev\~{a}o Prado contributed critically to the drafts and gave final approval for publication.
\printbibliography

@Article{berlow2004,
  author  = {Berlow, Eric L. and Neutel, Anje-Margiet and Cohen, Joel E. and De Ruiter, Peter C. and Ebenman, B.O. and Emmerson, Mark and Fox, Jeremy W. and Jansen, Vincent A. and Jones, J. Iwan and Kokkoris, Giorgos D. and others},
  journal = {Journal of Animal Ecology},
  title   = {Interaction strengths in food webs: issues and opportunities},
  year    = {2004},
  number  = {3},
  pages   = {585-598},
  volume  = {73}
}

@article{delattre2014note,
  title={A note on BIC in mixed-effects models},
  author={Delattre, Maud and Lavielle, Marc and Poursat, Marie-Anne},
  journal={Electronic journal of statistics},
  volume={8},
  number={1},
  pages={456--475},
  year={2014},
  publisher={Institute of Mathematical Statistics and Bernoulli Society}
}

@incollection{verdade2013counting,
  title={Counting Capybaras},
  author={Verdade, Luciano M. and Moreira, Jos{\'e} Roberto and Ferraz, Katia Maria P.M.B.},
  booktitle={Capybara},
  pages={357--370},
  year={2013},
  publisher={Springer}
}

@Article{yamaura2012biodiversity,
  title={Biodiversity of man-made open habitats in an underused country: a class of multispecies abundance models for count data},
  author={Yuichi Yamaura and J. Andrew Royle and Naoaki Shimada and Seigo Asanuma and Tamotsu Sato and Hisatomo Taki and Shun'ichi Makino},
  journal={Biodiversity and Conservation},
  volume={21},
  number={6},
  pages={1365--1380},
  year={2012},
  publisher={Springer}
}

@Article{golding2017multispecies,
  title={A multispecies dependent double-observer model: a new method for estimating multispecies abundance},
  author={Jessie D. Golding and J. Joshua Nowak and Victoria J. Dreitz },
  journal={Ecology and evolution},
  volume={7},
  number={10},
  pages={3425--3435},
  year={2017},
  publisher={Wiley Online Library}
}

@Proceedings{plummer2003jags,
  title        = {JAGS: A program for analysis of Bayesian graphical models using Gibbs sampling},
  year         = {2003},
  number       = {125.10},
  organization = {Vienna, Austria.},
  volume       = {124},
  author       = {Plummer, Martyn and others},
  booktitle    = {Proceedings of the 3rd international workshop on distributed statistical computing},
  pages        = {1--10},
}

@Manual{Plummer2017,
  title     = {JAGS Version 4.3.0 user manual},
  author    = {Plummer, Martyn},
  year      = {2017},
  publisher = {Lyon, France},
  url       = {http://www.stat.yale.edu/~jtc5/238/materials/jags_4.3.0_manual_with_distributions.pdf},
}

@Article{dorazio2005estimating,
  author    = {Robert M. Dorazio and J. Andrew Royle},
  journal   = {Journal of the American Statistical Association},
  title     = {Estimating size and composition of biological communities by modeling the occurrence of species},
  year      = {2005},
  number    = {470},
  pages     = {389--398},
  volume    = {100},
  publisher = {Taylor \& Francis},
}

@Article{king1972census,
  author    = {James G. King and Fred C. Robards and Calvin J. Lensink},
  journal   = {The Journal of Wildlife Management},
  title     = {Census of the bald eagle breeding population in southeast Alaska},
  year      = {1972},
  pages     = {1292--1295},
  publisher = {JSTOR},
}

@Article{hansen1987regulation,
  title={Regulation of bald eagle reproductive rates in southeast Alaska},
  author={Hansen, Andrew J.},
  journal={Ecology},
  volume={68},
  number={5},
  pages={1387--1392},
  year={1987},
  publisher={Wiley Online Library}
}

@Article{hodges2011bald,
  title={Bald Eagle population surveys of the north Pacific Ocean, 1967--2010},
  author={Hodges, John I.},
  journal={Northwestern Naturalist},
  volume={92},
  number={1},
  pages={7--12},
  year={2011},
  publisher={BioOne}
}

@Article{Gomez2017,
  author  = {Juan P. Gomez and Scott K. Robinson and Jason K. Blackburn and José M. Ponciano},
  journal = {Methods in Ecology and Evolution},
  title   = {An efficient extension of N-mixture models for multi-species abundance estimation},
  year    = {2017},
  month   = jul,
  number  = {2},
  pages   = {340-353},
  volume  = {9},
}

@Article{RoyleJ2004,
  author  = {J. Andrew Royle},
  journal = {Biometrics},
  title   = {N--mixture models for estimating population size from spatially replicated counts.},
  year    = {2004},
  number  = {1},
  pages   = {108-115},
  volume  = {60},
}

@Article{Foikanos2011,
  author  = {Konstantinos Fokianos and Dag Tjøstheim},
  journal = {Journal of Multivariate Analysis},
  title   = {Log-linear poisson autoregression},
  year    = {2011},
}

@Conference{Herdin2005,
  author    = {Markus Herdin and Nicolai Czink and Hüseyin Ozcelik and Ernst Bonek},
  booktitle = {2005 IEEE 61st Vehicular Technology Conference},
  title     = {Correlation matrix distance, a meaningful measure for evaluation of non-stationary MIMO Channels},
  year      = {2005},
  pages     = {146-140},
  volume    = {1},
}

@article{dennis2015computational,
  title={Computational aspects of N-mixture models},
  author={Dennis, Emily B. and Morgan, Byron J.T and Ridout, Martin S.},
  journal={Biometrics},
  volume={71},
  number={1},
  pages={237--246},
  year={2015},
  publisher={Wiley Online Library}
}

@Manual{RCT2020,
  title        = {R: A Language and Environment for Statistical Computing},
  address      = {Vienna, Austria},
  author       = {{R Core Team}},
  organization = {R Foundation for Statistical Computing},
  year         = {2020},
  url          = {https://www.R-project.org/},
}

@Manual{Su2020,
  title  = {R2jags: Using R to Run 'JAGS'},
  author = {Yu-Sung Su and Masanao Yajima,},
  note   = {R package version 0.6-1},
  year   = {2020},
  url    = {https://CRAN.R-project.org/package=R2jags},
}

@Article{lawrence1989concordance,
  author    = {Lin, Lawrence I-Kuei},
  journal   = {Biometrics},
  title     = {A concordance correlation coefficient to evaluate reproducibility},
  year      = {1989},
  pages     = {255--268},
  publisher = {JSTOR},
}

@article{todd1982food,
  title={Food habits of bald eagles in Maine},
  author={Todd, Charles S. and Young, L.S. and Owen Jr, Ray B. and Gramlich, Francis J.},
  journal={The Journal of Wildlife Management},
  pages={636--645},
  year={1982},
  publisher={JSTOR}
}

@article{dunstan1975food,
  title={Food habits of bald eagles in north-central Minnesota},
  author={Dunstan, Thomas C. and Harper, James F.},
  journal={The Journal of Wildlife Management},
  pages={140--143},
  year={1975},
  publisher={JSTOR}
}

@article{mcewan1980food,
  title={Food habits of the bald eagle in north-central Florida},
  author={McEwan, Linda C. and Hirth, David H.},
  journal={The Condor},
  volume={82},
  number={2},
  pages={229--231},
  year={1980},
  publisher={JSTOR}
}

@Misc{Pardieck2020,
  author       = {Pardieck, Keith L. and Ziolkowski, David and Michael Lutmerding and Veronica Aponte and Marie-Anne R. Hudson},
  howpublished = {U.S. Geological Survey data release},
  title        = {North American Breeding Bird Survey Dataset 1966-2019},
  year         = {2020},
  doi          = {https://doi.org/10.5066/P9J6QUF6},
}

@book{mccullagh1989generalized,
  author = {McCullagh, Peter and Nelder, John A.},
  publisher = {Chapman \& Hall},
  series = {Chapman and Hall/CRC Monographs on Statistics and Applied Probability Series},
  title = {Generalized Linear Models, Second Edition},
  year = 1989
}

@Article{witmer2005wildlife,
  title={Wildlife population monitoring: some practical considerations},
  author={Witmer, Gary W.},
  journal={Wildlife Research},
  volume={32},
  number={3},
  pages={259--263},
  year={2005},
  publisher={CSIRO Publishing}
}

@Article{nichols2004abundance,
  title={Abundance estimation and conservation biology},
  author={Nichols, James D. and MacKenzie, Darryl I.},
  journal={Animal biodiversity and conservation},
  volume={27},
  number={1},
  pages={437--439},
  year={2004}
}

@Article{yamaura2011modelling,
  title={Modelling community dynamics based on species-level abundance models from detection/nondetection data},
  author={Yuichi Yamaura and J. Andrew Royle and Kouji Kuboi and Tsuneo Tada and Susumu Ikeno and Shun'ichi Makino},
  journal={Journal of applied ecology},
  volume={48},
  number={1},
  pages={67--75},
  year={2011},
  publisher={Wiley Online Library}
}

@Article{royle2003estimating,
  title={Estimating abundance from repeated presence--absence data or point counts},
  author={Royle, J. Andrew and Nichols, James D.},
  journal={Ecology},
  volume={84},
  number={3},
  pages={777--790},
  year={2003},
  publisher={Wiley Online Library}
}

@article{dorazio2014estimating,
  title={Estimating abundances of interacting species using morphological traits, foraging guilds, and habitat},
  author={Dorazio, Robert M. and Connor, Edward F.},
  journal={PloS one},
  volume={9},
  number={4},
  pages={e94323},
  year={2014},
  publisher={Public Library of Science San Francisco, USA}
}

@Article{moral2018models,
  title={Models for Jointly Estimating Abundances of Two Unmarked Site-Associated Species Subject to Imperfect Detection},
  author={Moral, Rafael de Andrade and Hinde, John and Dem{\'e}trio, Clarice G.B. and Reigada, Carolina and Godoy, Wesley A.C.},
  journal={Journal of Agricultural, Biological and Environmental Statistics},
  volume={23},
  number={1},
  pages={20--38},
  year={2018},
  publisher={Springer}
}

@Article{gelman1992inference,
  title={Inference from iterative simulation using multiple sequences},
  author={Gelman, Andrew and Rubin, Donald B.},
  journal={Statistical science},
  volume={7},
  number={4},
  pages={457--472},
  year={1992},
  publisher={Institute of Mathematical Statistics}
}

@Article{alvarez2014bayesian,
  title={Bayesian inference for a covariance matrix},
  author={Alvarez, Ignacio and Niemi, Jarad and Simpson, Matt},
  journal={arXiv preprint arXiv:1408.4050},
  year={2014}
}

@Article{mann1945nonparametric,
  title={Nonparametric tests against trend},
  author={Mann, Henry B.},
  journal={Econometrica: Journal of the econometric society},
  pages={245--259},
  year={1945},
  publisher={JSTOR}
}

@Article{kendall1948rank,
  title={Rank correlation methods.},
  author={Kendall, Maurice George},
  year={1948},
  publisher={Griffin}
}

@Article{niku2019gllvm,
  title={gllvm: Fast analysis of multivariate abundance data with generalized linear latent variable models in r},
  author={Niku, Jenni and Hui, Francis K.C. and Taskinen, Sara and Warton, David I.},
  journal={Methods in Ecology and Evolution},
  volume={10},
  number={12},
  pages={2173--2182},
  year={2019},
  publisher={Wiley Online Library}
}

@Manual{Rsoftware,
    title = {R: A Language and Environment for Statistical Computing},
    author = {{R Core Team}},
    organization = {R Foundation for Statistical Computing},
    address = {Vienna, Austria},
    year = {2020},
    url = {https://www.R-project.org/},
  }

\appendix
\section{Analytic Correlations}\label{AppendixA}
In this section, we present the analytical expressions for the correlation between the latent abundances ($N_{s}$ and $N_{s'}$) for all $s \neq s'$ for the MNM model. For convenience of notation, we drop the dependence on $i$ and $t$,  $Y_{s} = (\{Y_{its}\})$, $N_{s} = (\{N_{is}\})$, $p_{s} = (\{p_{its}\})$, $\lambda_{s} = (\{\lambda_{is}\})$. We need the expectation, variance and covariance of the log-normally distributed $\lambda_{s}$, which are given by
\begin{linenomath*}
\postdisplaypenalty=0
\begin{align*}
   \mathbb{E}(\lambda_{s}) & = e^{\mu_{s} +\frac{1}{2} \Sigma_{ss}}, \\
    \mbox{Var}(\lambda_{s})& =e^{\mu_{s}+\mu_{s} +\frac{1}{2}(\Sigma_{ss}+\Sigma_{ss})}(e^{\Sigma_{ss}}-1) e^{2\mu_{s}+\Sigma_{ss}}(e^{\Sigma_{ss}}-1),\\
    \text{Cov}(\lambda_{s}, \lambda_{s'}) &= e^{\mu_{s} +\mu_{s'} +\frac{1}{2}(\Sigma_{ss}+\Sigma_{s's'})}(e^{\Sigma_{ss'}}-1).
\end{align*}
\end{linenomath*}
\noindent
where $\mu=\boldsymbol{\mu_{a}}+\textbf{x}_{i}\boldsymbol{\mu_{\beta}}$ and $\Sigma=\Sigma_{a}+\textbf{x}_{i}^{2}\boldsymbol{\Sigma_{\beta}}$. As $N_{s} \sim \text{Poisson}(\lambda_{s})$ and $Y_{s} \sim \text{Binomial}(N_{s}, p_{s})$, we can write the conditional expectation and variance directly:
\begin{linenomath*}
\postdisplaypenalty=0
\begin{align*}
\mathbb{E}(N_{s} \mid \lambda_{s}) &=\mbox{Var}(N_{s} \mid \lambda_{s})=\lambda_{s}, \\
\mbox{Var}(Y_{s} \mid N_{s}) &= N_{s}p_{s}(1-p_{s}).
\end{align*}
\end{linenomath*}
\noindent
The unconditional expectation and variance of $N_{s}$, and the unconditional covariance between $N_{s}$ and $N_{s'}$,  can be derived using the laws of total expectation, variance and covariance as follows:
\begin{linenomath*}
\postdisplaypenalty=0
\begin{align*}
\mathbb{E}(N_{s})&=\mathbb{E}_{\lambda_{s}}(\mathbb{E}_{N_{s}}(N_{s} \mid \lambda_{s})) = \mathbb{E}_{\lambda_{s}}(\lambda_{s})= e^{\mu_{s}+\frac{1}{2}\Sigma_{ss}}. \\
 \mbox{Var}(N_{s})&=\mathbb{E}_{\lambda_{s}}(\mbox{Var}_{N_{s}}(N_{s} \mid \lambda_{s}))+\mbox{Var}_{\lambda_{s}}(\mathbb{E}_{N_{s}}(N_{s} \mid \lambda_{s}))\\
&= \mathbb{E}_{\lambda_{s}}(\lambda_{s})\mbox{Var}_{\lambda_{s}}(\lambda_{s})= e^{\mu_{s}+\frac{1}{2}\Sigma_{ss}}+e^{2\mu_{s}+\Sigma_{ss}}(e^{\Sigma_{ss}}-1).\\
\text{Cov}(N_{s}, N_{s'})&=\mathbb{E}_{\lambda_{s}\lambda_{s'}}(\text{Cov}(N_{s}, N_{s'} \mid \lambda_{s}, \lambda_{s'})+\text{Cov}_{\lambda_{s}\lambda_{s'}}(\mathbb{E}_{Ns}(N_{s} \mid \lambda_{s}), \mathbb{E}_{N_{s'}}(N_{s'} \mid \lambda_{s'})).
\end{align*}
\end{linenomath*}
\noindent
 We assume that, given the correlated effects $\textbf{a}$, the latent and observed abundances are independent, which means that $\text{Cov}(N_{s}, N_{s'} \mid \lambda_{s}, \lambda_{s'})=0$. So, 
 \begin{linenomath*}
\postdisplaypenalty=0
\begin{align}
 \text{Cov}(N_{s}, N_{s'}) &= \text{Cov}_{\lambda_{s}\lambda_{s'}}(\mathbb{E}_{Ns}(N_{s} \mid \lambda_{s}), \mathbb{E}_{N_{s'}}(N_{s'} \mid \lambda_{s'})), \nonumber \\
 & =\text{Cov}(\lambda_{s}, \lambda_{s'}) = e^{\mu_{s} + \mu_{s'}+\frac{1}{2}(\Sigma_{ss}+\Sigma_{s's'})}(e^{\Sigma_{ss'}}-1),\nonumber\\
  &=(e^{\mu_{s}+\frac{1}{2}\Sigma_{ss}})(e^{\mu_{s'}+\frac{1}{2}\Sigma_{s's'}})(e^{\Sigma_{ss'}}-1) = \mathbb{E}(N_{s})\mathbb{E}(N_{s'})(e^{\Sigma_{ss'}}-1).\nonumber
\end{align}
\end{linenomath*}
\noindent
The correlations between $N_{s}$ and $N_{s'}$ can be estimated in a similar way for the other models presented in this paper. The models with a hurdle component require the use of Hurdle-Poisson  $\mathbb{E}(N_{s} \mid \lambda_{s})$ and $\text{Var}(N_{s} \mid \lambda_{s})$, which results in the need for an approximation of $\mathbb{E}(N_{s})$ and $\text{Var}(N_{s})$ and $\text{Cov}(N_{s}, N_{s'})$ based on quadratic Taylor expansions. Correlations for models with an autoregressive component follow the same form as the MNM model, with the following substitution for $\lambda$
\begin{linenomath*}
\postdisplaypenalty=0
\begin{align*}
 \lambda_{s} \sim \text{MVLN}(e^{\mu_{i}+\frac{1}{2}\Sigma_{ii}}, (e^{\mu_{i}+\mu_{j}+\frac{1}{2}(\Sigma_{ii}+\Sigma_{jj})})(e^{\Sigma_{ij}}-1)),
\end{align*}
\end{linenomath*}
where $\mu = \mu_{a} +\textbf{x}_{i}\mu_{\beta}+ \text{log}(N_{i,k-1,s}+1)\mu_{\phi}$ and $\Sigma = \Sigma_{a}+\textbf{x}_{i}^{2}\Sigma_{\beta}+\text{log}(N_{i,k-1,s}+1)^2\Sigma_{\phi}$.

\section{Simulation Study}\label{AppendixB}
In this section, we describe the simulation studies which were used to determine the accuracy of the estimates produced by the multi-species N-mixture models.

To determine if our modelling framework produces accurate estimates at contrasting sample sizes, a series of simulations were run in which we varied the number of sites, $R \in \{10,100\}$, the number of sampling occasions, $T \in \{5,10\}$, and the number of species observed, $S \in \{5,10\}$. Within these simulations, we varied the detection probability $p$, and the mean number of individuals per site $\lambda$. Small values for $p$ lay between 0.1 and 0.4, while large values for $p$ lay between 0.5 and 0.9. Small values for $\lambda$ had a median value of 7 and standard deviation of 10, while large values for $\lambda$ had a median value of 55 and standard deviation of 74.

In the case of the Hurdle and Hurdle-AR models, we also varied the probability of a zero-count occurring, $\theta \in \{0.2, 0.7\}$. For each combination of parameters, we simulated 100 datasets and estimated $N$, $\boldsymbol{\Sigma}_{a}$, and $p$. We also estimated values for $\theta$ and $\phi$, in the case of the Hurdle and AR models, respectively. Relative mean bias was calculated for the estimated probability of obtaining a zero count $\hat{\theta}$, autocorrelation coefficient $\hat{\phi}$, probability of detection $p$, and mean of the abundance random effects $\boldsymbol{\mu}_a$. The smaller the value for relative bias, the closer to the true value our estimated parameters were. We compared $\hat{N}$ to $N$ using the concordance correlation coefficient \parencite{lawrence1989concordance}, which is given by the formula: 
\begin{linenomath*}
\postdisplaypenalty=0
\begin{align}
  \rho_{c} = \frac{2 \rho \hat{\sigma} \sigma}{\hat{\sigma^{2}}+\sigma^{2}+(\hat{\mu}-\mu)^{2}}, \nonumber
\end{align}
\end{linenomath*}
where $\rho$ is Pearson's correlation coefficient, $\sigma$ and $\mu$ are the standard deviation and mean of the true values of $N$, and $\hat{\sigma}$ and $\hat{\mu}$ are the standard deviation and mean of the estimated values of $N$. The Pearson correlation coefficient is a measure of the strength of a linear association between two variables. However, the Pearson correlation is invariant under changes in location and scale. If two variables exhibit a linear relationship, but are very different in terms of their location or scale, the Pearson correlation coefficient will not reveal this. The concordance correlation coefficient, however, does take into account differences in location and scale. For this reason, the concordance correlation coefficient was chosen as a measure of the linear relationship between the true abundance and estimated abundance, rather than the Pearson correlation coefficient. The higher the value of the concordance correlation coefficient, the closer our estimates for $N$ were to the true values. 

We compared our estimated correlation matrix to the true value using the correlation matrix distance \parencite{Herdin2005}, which is given by the following formula:
\begin{linenomath*}
\postdisplaypenalty=0
\begin{align}
   \text{CMD}(\mathbf{X_1},\mathbf{X_2}) = 1-\frac{\text{tr}(\mathbf{X_1}\mathbf{X_2})}{\norm{\mathbf{X_1}}_{f}\norm{\mathbf{X_2}}_{f}}, \nonumber
\end{align}
\end{linenomath*}
where $\mathbf{X_1}$ and $\mathbf{X_2}$ are two correlation matrices, $\text{tr}(\mathbf{X_1}\mathbf{X_2})$ is the trace of the product of these two matrices, and $\norm{.}_{f}$ denotes the Frobenius norm. 

Additionally, the coverage probabilities for each parameter were determined as the proportion of simulations in which the 50\% credible interval contained the true parameter value. We expect that approximately 50\% of the time, the estimated 50\% credible interval for the parameter will contain the true value of that parameter (Appendix \ref{AppendixD}). Each of these scenarios were simulated 100 times. All data was simulated using the R statistical software version 4.0.2 \parencite{RCT2020}, and all Bayesian models were implemented using the \texttt{R2jags} package \parencite{Su2020}.

\subsection{Simulation Study Results}
The results of the small-scale simulation study, which was composed of data simulated for five species at 10 sites, over five years, is shown in Table \ref{table:small}. The results of the large-scale simulation study, which contained 10 species, 100 sites and 10 years, is shown in Table \ref{table:large}.

\subsubsection{MNM Model}
The large-scale simulation study (Table \ref{table:large}) produced reliable estimates for latent abundance $N$ at every combination of $p$ and $\lambda$, with CCC values between 0.97 and 0.99. Estimates of $N$ from the small-scale simulation study (Table \ref{table:small}) appear more dependent on the detection probability $p$, with greater CCC values associated with larger detection probabilities. 

From  Table \ref{table:large}, the relative bias for the estimate of $p$ shows that when $R$, $T$ and $S$ are large, the model produces estimates for $p$ which are accurate to two decimal places. When, $R$, $T$ and $S$ are small (Table \ref{table:small}), the relative bias for the estimate of $p$ is larger for small median $p$. When $R$, $T$ and $S$ are small, larger values of $p$ produce more reliable estimates of $p$.

Estimates for the correlation matrix and $\boldsymbol{\mu}_{a}$ improve with larger values of $\lambda$. In both Table \ref{table:small} and Table \ref{table:large}, the relative bias for $\boldsymbol{\mu}_{a}$ and the CMD decrease when $\lambda$ is larger. Larger values of $R$, $T$ and $S$ produce more accurate estimates of the inter-species correlations and $\boldsymbol{\mu}_{a}$, as can be seen by the decrease in the sizes of the CMD and  RB($\boldsymbol{\mu}_{a}$) between Table \ref{table:small} and Table \ref{table:large}. 

Coverage probabilities (Appendix \ref{AppendixD}) for this model reveal that both small- and large-scale simulations produce parameters whose true value lie within the 50\% credible interval approximately 50\% of the time, as expected.
\subsubsection{Autoregressive Model}
At both small- (Table \ref{table:small}) and large-scale simulations (Table \ref{table:large}), the autoregressive model produced reliable estimates for $N$, with CCC values above 0.9 for all simulations. Both the Table \ref{table:small} and Table \ref{table:large} see CMD values accurate to two decimal places. Relative bias for $p$ decreases as median $p$ increases. This can be seen for both small (Table \ref{table:small}) and large (Table \ref{table:large}) values of $R$, $T$, and $S$. In Table \ref{table:small}, relative bias for the autocorrelation coefficient $\phi$ is much larger for small abundance. In this situation, the estimates for the autocorrelation coefficient $\phi$ cannot be relied upon. This is an issue that persists, though not as severely, as $R$, $T$ and $S$ increase in Table \ref{table:large}.

All parameters in this model have coverage probabilities of approximately 50\%, as is expected for the 50\% credible intervals (Appendix \ref{AppendixD}).
{
\renewcommand{\baselinestretch}{1.2}
\begin{table}[H]
\begin{tabular}{c c c c c c c c c}
 \hline
 Median $p$  & Median $\lambda$ & $\theta$ & CCC & CMD & RB($p$)& RB($\boldsymbol{\mu}_{a}$) & RB($\theta$) & RB($\phi$)\\
 \hline
\multicolumn{1}{l}{MNM}\\
0.3 &	7&  -  & 0.7871&	0.1037&	0.2512&0.1269 & - & - \\
0.3&	55& - &	0.8689&	0.0641&	0.2256& 0.0554 & - & - \\
0.8&	7& - &	0.9847&	0.0768&	0.0612& 0.0379& - & - \\
0.8&	55& - &	0.9878&	0.0522&	0.0579&0.013 & - & - \\
\multicolumn{1}{l}{Hurdle}\\
0.3&	7&	0.2&	0.772&	0.146&	0.437&0.2499	&0.241 & - \\
0.3&	7&	0.7&	0.808&	0.196&	0.418&0.3485	&0.116 & - \\
0.3&	55&	0.2&	0.799&	0.081&	0.418&0.1065	&0.191 & - \\
0.3&	55&	0.7&	0.816&	0.152&	0.469&0.1194	&0.134 & - \\
0.8&	7&	0.2&	0.96&	0.114&	0.071&0.1385	&0.242 & - \\
0.8&	7&	0.7&	0.928&	0.167&	0.094&0.1923	&0.13 & - \\
0.8&	55&	0.2&	0.927&	0.074&	0.085&0.0712	&0.201 & - \\
0.8&	55&	0.7&	0.946&	0.125&	0.103&0.0811	&0.114 & - \\
\multicolumn{1}{l}{AR}\\
0.3&	7& - &	0.9475&	0.0638&	0.1529& 0.14 & - &	1.9807\\
0.3&	55& - &	0.9818&	0.0515&	0.1479&0.0663 & - &	0.0515\\
0.8&	7& - &	0.999&	0.0652&	0.0153&0.1149 & - &	1.8982\\
0.8&	55& - &	0.9999&	0.0598&	0.0168& 0.0538 & - &	0.0598\\
\multicolumn{1}{l}{Hurdle-AR}\\
0.3&	7&	0.2& 0.9227&	0.0602&	0.1919&0.1475	&0.1145 &  2.758\\
0.3&	7&	0.7& 0.8899&	0.1262&	0.3228&0.2084	&0.0329&  12.365\\
0.3&	55&	0.2& 0.9716&	0.0709&	0.1866&0.0752	&0.0959 &  0.649\\
0.3&	55&	0.7& 0.9223&	0.0918&	0.3575&0.0941	&0.0356 &  4.206\\
0.8&	7&	0.2& 0.994&	    0.0718&	0.0386&	0.1183&0.1024 &  2.088\\
0.8&	7&	0.7& 0.974&	    0.0972&	0.0787&0.1457	&0.0316 &  6.459\\
0.8&	55&	0.2& 0.9977&	0.0636&	0.0355&	0.0497&0.1112 &  0.419\\
0.8&	55&	0.7& 0.9322&	0.1101&	0.0986&	0.0775&0.0308 &  5.505\\
 \hline
\end{tabular}
\caption{Results of small-scale simulation ($(R,T,S,K)=(10,5,5,5)$): Concordance Correlation Coefficient (CCC) for the estimates of latent abundance $N$, Correlation Matrix Distance (CMD) for the estimate of the inter-species correlations, and relative biases for probability of detection ($p$), probability of obtaining a zero count ($\theta$), and autocorrelation coefficient ($\phi$).}
\label{table:small}
\end{table}}

{
\renewcommand{\baselinestretch}{1.2}
\begin{table}[H]
\begin{tabular}{c c c c c c c c c}
 \hline
 Median $p$  & Median $\lambda$ & $\theta$ & CCC & CMD & RB($p$) &  RB($\boldsymbol{\mu}_{a}$)&RB($\theta$) & RB($\phi$)\\
 \hline
\multicolumn{1}{l}{MNM}\\
0.3 &	7& - &	0.9735&	0.0315&	0.0661& 0.0163& - & - \\
0.3&	55& - &	0.9902&	0.0197&	0.0613& 0.01 & - & - \\
0.8&	7& - &	0.9972&	0.026&	0.0162& 0.0104& - & - \\
0.8&	55& - &	0.999&	0.0176&	0.0171&0.0041 & - & - \\
\multicolumn{1}{l}{Hurdle}\\
0.3&	7&	0.2&	0.953&	0.049&	0.183&0.0881	&0.058 & - \\
0.3&	7&	0.7&	0.948&	0.193&	0.278&0.131	&0.016 & - \\
0.3&	55&	0.2&	0.964&	0.022&	0.192&0.042	&0.061 & - \\
0.3&	55&	0.7&	0.928&	0.152&	0.462&0.0879	&0.015 & - \\
0.8&	7&	0.2&	0.997&	0.041&	0.019&0.0389	&0.058 & - \\
0.8&	7&	0.7&	0.997&	0.179&	0.033&0.0569	&0.015 & - \\
0.8&	55&	0.2&	0.999&	0.023&	0.024&0.018	&0.046 & - \\
0.8&	55&	0.7&	0.999&	0.206&	0.034&0.0546	&0.013 & - \\
\multicolumn{1}{l}{AR}\\
0.3&	7& - &	0.9923&	0.0174&	0.05& 0.0443 &- &	0.2799\\
0.3&	55& - &	0.9971&	0.0182&	0.0469& 0.0192&- &	0.0802\\
0.8&	7& - &	0.9997&	0.0182&	0.0044& 0.032 &- &	0.2433\\
0.8&	55& -  & 0.9999&	0.0204&	0.0054& 0.016 &- 	&0.0704\\
\multicolumn{1}{l}{Hurdle-AR}\\
0.3&	7&	0.2& 0.9937&	0.02&	0.0567&0.0417 & 0.0248 & 0.269 \\
0.3&	7&	0.7& 0.9892&	0.0257&	0.0679&0.0535 & 0.0109& 0.434\\
0.3&	55&	0.2& 0.9779&	0.0143&	0.0662&0.0223 & 0.0276 & 0.069\\
0.3&	55&	0.7& 0.9678&	0.0168&	0.0943&0.0258& 0.0128 & 0.109 \\
0.8&	7&	0.2& 0.9994&	0.0178&	0.0078&0.0335 &0.0227 & 0.315\\
0.8&	7&	0.7& 0.9992&	0.0241&	0.0096&	0.0371 & 0.0123  & 0.396\\
0.8&	55&	0.2& 0.9967&	0.0187&	0.0136& 0.0158 & 0.0245 & 0.049\\
0.8&	55&	0.7& 0.9976&	0.0166&	0.0179&0.0180 &0.0179 & 0.112\\
 \hline
\end{tabular}
\caption{Results of large-scale simulation ($(R,T,S,K)=(100,10,10,10)$): Concordance Correlation Coefficient (CCC) for the estimates of latent abundance $N$, Correlation Matrix Distance (CMD) for the estimate of the inter-species correlations, and relative biases for probability of detection ($p$), probability of obtaining a zero count ($\theta$), and autocorrelation coefficient ($\phi$).}
\label{table:large}
\end{table}}

\subsubsection{Hurdle Model}
Similar to the MNM model, when $R$, $T$ and $S$ are large (Table \ref{table:large}), consistently accurate estimates of latent abundance $N$ are produced, with CCC values between 0.948 and 0.999.  In Table \ref{table:small} we see that CCC values depend more on the detection probability, with more accurate estimates of $N$ produced when detection probability is high. Both Table \ref{table:small} and Table \ref{table:large} show higher accuracy in estimates of the inter-species correlations when zero-inflation is small, and abundance is large. CMD values are greater when $\theta = 0.7$ or  median $\lambda = 7$ than for $\theta = 0.2$ or median $\lambda = 55$. From both Table \ref{table:small} and Table \ref{table:large}, the Hurdle model sees much smaller relative bias for $p$ when median $p$ is large compared to when median $p$ is small. Relative bias for $\theta$ decreases when $\theta$ increases, indicating that $\theta$ is estimated with more accuracy when zero-inflation is large. Table \ref{table:large} sees smaller relative bias for $\theta$ than Table \ref{table:small}, revealing that the strength of zero-inflation $\theta$ is estimated more accurately when $R$, $T$ and $S$ are large.

Issues with parameter convergence were encountered when fitting the Hurdle model. When zero-inflation and abundance are large, and detection probability is small, issues with convergence occurred in up to 20$\%$ of parameters. While this convergence issue does not appear to negatively affect the relative biases of parameter estimates, as can be seen in Table \ref{table:small} and Table \ref{table:large}, coverage probability for detection probability $p$ and random effect mean $\mu_{a}$ is negatively impacted (Appendix \ref{AppendixD}). We also see coverage for $N$ which is larger than 50\%. This is to be expected, and due to zero counts being perfectly predicted.

\subsubsection{Hurdle-Autoregressive Model}
In Table \ref{table:small}, CCC values demonstrate that the Hurdle-AR model produces estimates for $N$ which are more accurate when the probability of obtaining a zero count is smaller. However, increasing $R$, $T$, and $S$ (Table \ref{table:large}) reduces this dependence on $\theta$, and all CCC values produced are greater than 0.95.

The small-scale simulation (Table \ref{table:small}) has CMD values and relative biases for $p$ and $\boldsymbol{\mu}_{a}$ which increase when the probability of obtaining a zero count increases, indicating that the inter-species correlations, $p$ and $\boldsymbol{\mu}_{a}$ are estimated more accurately when the degree of zero-inflation is low. The same is true for the large-scale simulation (Table \ref{table:large}), though the differences in CMD and relative biases between small $\theta$ and large $\theta$ are not as large, revealing that the increase in $R$, $T$ and $S$ renders the increase in zero-inflation less important in the estimation of these parameters.

The Hurdle-AR model suffers with the same issue estimating $\phi$ when the probability of obtaining a zero count is high. This issue is more severe in Table \ref{table:small}, and estimates of $\phi$ cannot be trusted when $R$, $T$ and $S$ are small but $\theta$ is large. Like the AR model, this issue is not as acute in Table \ref{table:large}, as an increase in $R$, $T$ and $S$ appears to compensate for the problems caused by large zero-inflation.

\section{Estimated Abundances}\label{AppendixC}
In this section, we provide a comparison of the maximum observed abundance with the maximum abundance estimated from the Hurdle(B) Model, fitted to the NABBS data.
\begin{table}[H]
    \centering
    \begin{tabular}{ccc}
    \hline
        Species  & Max. Y & Max. N \\
        \hline
        Bald Eagle &	29&	39\\
        Canada Goose&	24&	46\\
        Hammond's Flycatcher&	4&	10\\
        Red-breasted Sapsucker&	4&	22\\
        Steller's Jay&	6&	14\\
        Swainson's Thrush&	12&	71\\
        Tree Swallow&	50&	53\\
        Trumpeter Swan&	38&	39\\
        Varied Thrush&	21&	70\\
        Wilson's Snipe&	6&	43\\
       \hline
    \end{tabular}
    \caption{Maximum observed and estimated abundances per species, produced by the Hurdle(B) model.}
    \label{tab:maxN}
\end{table}

\section{Coverage Probabilities} 
\label{AppendixD}

{
\renewcommand{\baselinestretch}{1.2}
\begin{table}[H]
\centering
\begin{tabular}{c c c c c c c c c}
 \hline
 \multicolumn{3}{c}{True Median Value} & \multicolumn{6}{c}{Coverage}\\
 \cmidrule(lr){1-3}\cmidrule(lr){4-9}
$p$  & $\lambda$ & $\theta$ & N & $\Sigma$ & $p$ & $\mu_{a}$ & $\theta$ & $\phi$\\
 \hline
\multicolumn{1}{l}{MNM}\\
0.3 &	7&  -  & 0.55 & 0.53 &0.49&0.49 & - & -\\
0.3&	55& - & 0.49 & 0.51 & 0.48 & 0.45 & - & -\\
0.8&	7& - & 0.56 & 0.52 & 0.47 & 0.54 & - & -\\
0.8&	55& - & 0.52 & 0.52 & 0.51 & 0.58 & - & -\\
\multicolumn{1}{l}{Hurdle}\\
0.3&	7&	0.2& 0.61 & 0.57 &0.45& 0.52 & 0.54 & -\\
0.3&	7&	0.7& 0.72 &0.60 & 0.47 &0.50 & 0.6 & -\\
0.3&	55&	0.2& 0.59 & 0.51& 0.49 & 0.51 & 0.45 & -\\
0.3&	55&	0.7& 0.70 & 0.52& 0.51 &0.52 & 0.50 & -\\
0.8&	7&	0.2& 0.64 &0.51 & 0.53 &0.51 & 0.42 & -\\
0.8&	7&	0.7& 0.73& 0.55 &0.53&0.55 & 0.54 & -\\
0.8&	55&	0.2& 0.60 &0.52& 0.52 &0.52 & 0.53 & -\\
0.8&	55&	0.7& 0.69& 0.56& 0.46 &0.56 & 0.54 & -\\
\multicolumn{1}{l}{AR}\\
0.3&	7& - & 0.59 & 0.53 & 0.54 & 0.52 & - & 0.52\\
0.3&	55& - & 0.53 & 0.54 & 0.53 & 0.49 & - & 0.45\\
0.8&	7& - & 0.54 & 0.51 & 0.52 & 0.48 & - & 0.50\\
0.8&	55& - &	0.52 & 0.52 &0.51 & 0.49 & - & 0.47\\
\multicolumn{1}{l}{Hurdle-AR}\\
0.3&	7&	0.2& 0.66 & 0.54 & 0.54 & 0.56 & 0.42 & 0.47\\
0.3&	7&	0.7& 0.85 & 0.53 & 0.51 & 0.56 & 0.46 & 0.46\\
0.3&	55&	0.2& 0.62 & 0.49 & 0.50 & 0.51 & 0.52 & 0.49\\
0.3&	55&	0.7& 0.84 & 0.51 & 0.52 & 0.50 & 0.48 & 0.50\\
0.8&	7&	0.2& 0.66 & 0.52 & 0.55 & 0.50 & 0.54 & 0.51\\
0.8&	7&	0.7& 0.87 & 0.51 & 0.46 & 0.47 & 0.54 & 0.50\\
0.8&	55&	0.2& 0.63 & 0.52 & 0.58 & 0.49 & 0.48 & 0.48\\
0.8&	55&	0.7& 0.84 & 0.52 & 0.47 & 0.50 & 0.54 & 0.41\\
 \hline
\end{tabular}
\caption{Proportion of small-scale simulations ($(R,T,S,K)=(10,5,5,5)$) in which the true parameter value lies within the estimated 50\% credible interval.}
\label{table:coverageprob1}
\end{table}}

{
\renewcommand{\baselinestretch}{1.2}
\begin{table}[H]
\centering
\begin{tabular}{c c c c c c c c c}
 \hline
 \multicolumn{3}{c}{True Median Value} & \multicolumn{6}{c}{Coverage}\\
 \cmidrule(lr){1-3}\cmidrule(lr){4-9}
$p$  & $\lambda$ & $\theta$ & N & $\Sigma$ & $p$ & $\mu_{a}$ & $\theta$ & $\phi$\\
 \hline
\multicolumn{1}{l}{MNM}\\
0.3 &	7&  -  & 0.55 & 0.53 &0.46&0.50 & - & -\\
0.3&	55& - & 0.49 & 0.52 & 0.45 & 0.47 & - & -\\
0.8&	7& - & 0.52 & 0.52 & 0.49 & 0.43 & - & -\\
0.8&	55& - & 0.51 & 0.53 & 0.50 & 0.42 & - & -\\
\multicolumn{1}{l}{Hurdle}\\
0.3&	7&	0.2& 0.61 & 0.53 &0.38& 0.39 & 0.43 & -\\
0.3&	7&	0.7& 0.84 &0.54 & 0.37 &0.38 & 0.51 & -\\
0.3&	55&	0.2& 0.51 & 0.54& 0.36 & 0.36 & 0.46 & -\\
0.3&	55&	0.7& 0.78 & 0.55& 0.30 &0.25 & 0.54 & -\\
0.8&	7&	0.2& 0.62 &0.51 & 0.53 &0.52 & 0.44 & -\\
0.8&	7&	0.7& 0.85& 0.53 &0.51&0.51 & 0.57 & -\\
0.8&	55&	0.2& 0.61 &0.55& 0.49 &0.48 & 0.58 & -\\
0.8&	55&	0.7& 0.85& 0.56& 0.50 &0.41 & 0.6\\
\multicolumn{1}{l}{AR}\\
0.3&	7& - & 0.58 & 0.56 & 0.48 & 0.45 & - & 0.48\\
0.3&	55& - & 0.52 & 0.50 & 0.51 & 0.52 & - & 0.51\\
0.8&	7& - & 0.52 & 0.53 & 0.48 & 0.51 & - & 0.52\\
0.8&	55& - &	0.49 & 0.47 &0.53 & 0.47 & - & 0.46\\
\multicolumn{1}{l}{Hurdle-AR}\\
0.3&	7&	0.2& 0.66 & 0.53 & 0.46 & 0.53 & 0.45 & 0.50\\
0.3&	7&	0.7& 0.78 & 0.51 & 0.50 & 0.45 & 0.56 & 0.52\\
0.3&	55&	0.2& 0.60 & 0.56 & 0.49 & 0.49 & 0.40 & 0.48\\
0.3&	55&	0.7& 0.74 & 0.54 & 0.47 & 0.52 & 0.40 & 0.53\\
0.8&	7&	0.2& 0.63 & 0.56 & 0.47 & 0.48 & 0.48 & 0.45\\
0.8&	7&	0.7& 0.76 & 0.53 & 0.48 & 0.47 & 0.44 & 0.50\\
0.8&	55&	0.2& 0.59 & 0.54 & 0.49 & 0.49 & 0.42 & 0.49\\
0.8&	55&	0.7& 0.74 & 0.55 & 0.41 & 0.44 & 0.40 & 0.50\\
 \hline
\end{tabular}
\caption{Proportion of large-scale simulations ($(R,T,S, K)=(100,10,10,10)$) in which the true parameter value lies within the estimated 50\% credible interval.}
\label{table:coverageprob2}
\end{table}}
\end{document}